# Comment: Struggles with Survey Weighting and Regression Modeling

**Sharon L. Lohr**

In the ideal samples of survey sampling textbooks, weights are the inverses of the inclusion probabilities for the units. But nonresponse and undercoverage occur, and survey statisticians try to compensate for the resulting bias by adjusting the sampling weights. There has been much debate about when and whether weights should be used in analyses, and how they should be constructed. Professor Gelman deserves thanks for clarifying the discussion about weights and for raising interesting issues and questions.

If we use weights in estimation, what would we like them to accomplish? Here are some desirable properties:

1. The mean squared error (MSE) of estimators is smaller if the weights are used than if the weights are not used.
2. Estimators produced using the weights are internally consistent. Thus, if $\hat{Y}_1$ is the estimated total medical expense for men in the population, $\hat{Y}_2$ is the estimated total medical expense for women in the population and $\hat{Y}_3$ is the estimated total medical expense for everyone in the population, then $\hat{Y}_1 + \hat{Y}_2 = \hat{Y}_3$.
3. We may have independent population counts from a census or administrative data source for sex, age, race/ethnicity and other variables. If we apply the weights to estimate these quantities, the estimates equal the true population counts. We refer to this as the calibration property.

*Sharon L. Lohr is Thompson Industries Dean's Distinguished Professor of Statistics, Department of Mathematics and Statistics, Arizona State University, Tempe, Arizona 85287-1804, USA e-mail: sharon.lohr@asu.edu.*

4. The weight for unit $i$ in the sample can be thought of as the number of population units represented by unit $i$.
5. The estimators have optimal properties under superpopulation models that are thought to fit the data.
6. The estimators are robust to misspecifications of the superpopulation models.
7. The procedure for constructing the weights is objective and transparent.

All of these are good properties. The problem is that one can only rarely construct a set of weights that satisfies all of them simultaneously.

In this discussion, we distinguish between design weights and weighting adjustments used for poststratification. The design weights are

$$d_i = \frac{1}{P(\text{unit } i \text{ included in sample})}.$$

The design weight $d_i$ is a property of unit $i$; under design-based inference, it is a fixed constant. If two samples are drawn independently using the same probability sampling design and if each sample includes unit $i$, the weight $d_i$ for the unit is the same in each sample. Poststratification weight adjustments, however, depend on the selected sample $\mathcal{S}$. In the simplest case of ratio adjustment, we multiply each sampling weight $d_i$ by the factor $g_i(\mathcal{S}, x) = X/\hat{X}$, where $X$ is the known population total of auxiliary variable $x$ and $\hat{X} = \sum_{i \in \mathcal{S}} d_i x_i$. The resulting weight is $w_i(\mathcal{S}, x) = d_i g_i(\mathcal{S}, x)$; the weight depends on the sample selected and on the auxiliary variable $x$ through the estimated total $\hat{X}$. The ratio estimator of the population total is then $\hat{Y}_r = (X/\hat{X}) \sum_{i \in \mathcal{S}} d_i y_i = \sum_{i \in \mathcal{S}} w_i(\mathcal{S}, x) y_i$. Similarly, for generalized regression estimation,

$$g_i(\mathcal{S}, \mathbf{x}) = 1 + (\mathbf{X} - \hat{\mathbf{X}})^T \left( \sum_{j \in \mathcal{S}} d_j \mathbf{x}_j \mathbf{x}_j^T / c_j \right)^{-1} \mathbf{x}_i / c_i,$$

where the scaling constant $c_i$ may depend on $\mathbf{x}$. For the special case of poststratification, $g_i(\mathcal{S}, \mathbf{x}) = N_c / \hat{N}_c$ for observation $i$ in poststratification class $c$.

Thus, for poststratification, the weight adjustments are positive; for general regression models, however, the weight adjustments are unrestricted. The weight $w_i(\mathcal{S}, x)$ varies from sample to sample. Since the weight adjustment depends only on $x$, though, and not on $y$, the weight $w_i(\mathcal{S}, x)$ will be the same for every response variable used in that sample.

The weights proposed by Gelman using hierarchical models add dependence on $y$ to the mix. When a proper prior is used for $\boldsymbol{\beta}$ and the parameters in the covariance matrix are estimated from the data, the weights change if a different sample is taken and they also change if a different response variable is used; we denote this dependence on the response variable by expressing the weights as $w_i(\mathcal{S}, x, y)$.

Now let us look at how various weighting schemes address the properties listed above.

**1. Reducing the mean squared error.** The rationale for any kind of weighting adjustments is that estimators constructed using the weights should have smaller MSE than estimators constructed without using the weights. If the response variable has the same mean in each of the poststrata, then poststratification weights decrease efficiency since they increase the variance of the estimator (Kish, 1992). Korn and Graubard (1999, Chapter 4) show that even a few weights that differ greatly from the others can substantially increase the variance. But if the poststrata have different means, weights often decrease the bias of the estimator, particularly when there is substantial undercoverage or nonresponse. If there is a strong relationship between the response variable $y$ and the auxiliary information $\mathbf{x}$, then using poststratification or regression modeling to adjust the weights can also decrease the variance; in the best possible case with $y$ proportional to $x$, $\sum_{i \in \mathcal{S}} w_i(\mathcal{S}, \mathbf{x}) y_i = Y$ which has variance 0. One reason to poststratify is to try to compensate for nonresponse or undercoverage in certain poststrata so that potential biases are reduced. The population may have 100,000 urban residents and 30,000 rural residents. Because of nonresponse, a simple random sample may end up with 1100 urban residents and 200 rural residents, so that urban residents are overrepresented in the sample. If the urban and rural areas have different means, an unweighted estimator is biased for estimating the population mean.

In hopes of reducing the variance due to disparities in the weights and reducing the influence of observations with large weights, various researchers have proposed shrinkage methods for the weights. Smoothed weights also help protect the confidentiality of the data. Traditionally, statistical agencies have collapsed poststrata, or trimmed weight adjustments $g_i$ that are too large. The resulting weights depend on $\mathcal{S}$ and $x$, and usually do not depend on $y$, but are difficult to justify from an optimality perspective and have an ad hoc quality that some find disturbing.

Stokes (1990) shrinks the weights using an empirical Bayes approach. Elliott and Little (2000) shrink the weights using mixed models. Gelman's procedure smooths the weights by using a hierarchical model to shrink the regression parameter estimates. These procedures have desirable properties under the models used, but give weights that depend on $y$ as well as $\mathcal{S}$ and $\mathbf{x}$. Thus, a different set of weights would be used for each response variable.

**2. Internal consistency.** To have internal consistency, the weights need to be the same for each response variable. Weights of the form $w_i(\mathcal{S}, x, y)$ that depend on the response variable, such as those resulting from the hierarchical regression approach of Gelman's paper, can, as he points out, lead to internally inconsistent estimators. In addition, multivariate statistics are affected if different weights are used for different variables.

Alexander (1991), discussing papers on whether to use weights in regression models based on survey data, asked: "Are we really to use weighted results for some parts of a report and unweighted results for others?" A similar question can be asked here: Are we really to use one set of weights to estimate unemployment, another set of weights to estimate poverty, and yet another set to estimate the relation between poverty and unemployment? I think that for official statistics, internal consistency is very important and therefore weights that do not depend on $y$ are preferred.

One possibility for obtaining internal consistency is to obtain one set of shrinkage weights that is then used for all variables. Chambers (1996) and Rao and Singh (1997) proposed using ridge-regression methods to shrink the weights. These methods depend only on the $\mathbf{x}$ variables and not on $y$.

**3. Calibration.** Often calibration is desirable so that demographic counts and other quantities will be consistent across surveys, and be consistent with the census. Poststratified weights satisfy the calibration property. Shrinkage weights, in general, do not, although some methods are closer to satisfying it than others.

Calibration is often deemed more important for larger classifications than for finer classifications. For example, it might be considered important for the weighted counts to equal the population counts for sex × ethnicity categories, but less important for sex × ethnicity × age categories. Statistical agencies implicitly order the importance of calibration classifications when they devise procedures for collapsing poststratification cells.

In regression models for weights, exact calibration may be achieved by using a noninformative prior for all components of $\boldsymbol{\beta}$, that is, using classical regression. One possible variation of using hierarchical regression for constructing weights might be to use the prior distribution of $\boldsymbol{\beta}$ to try to control the degree of calibration for different variables. However, as soon as an informative prior is introduced for any component of $\boldsymbol{\beta}$, the calibration for the other variables can disappear. More research is needed on how the weights can be smoothed yet calibrate to the most important population quantities.

**4. Weight as population units represented by sample unit.** The design weights $d_i$ are commonly thought of as the number of population units represented by unit $i$ in the sample. The weights $w_i(\mathcal{S}, x)$ from poststratification can be thought of in the same way, as long as the adjustment is not too extreme. In some cases, though, the adjusted weight can be less than 1, which would lead to the interpretation that the sampled unit represents only a fraction of a unit in the population—that is, the sampled unit does not even represent itself.

Deville and Särndal (1992) point out that weights from generalized regression estimators can be negative, which presents even more problems for interpretation. Weights can be negative in the regression models in (7) and (9) of Gelman's paper when some interactions are omitted from the model. Thus if the regression model used to construct the weights has main effects terms for sex and ethnicity but does not contain the interaction term, it is easy to construct examples in which the weights of some observations are negative even though weights for females sum to the population count for females and the weights for blacks sum to the population count for blacks. Negative weights can be awkward to explain and are unacceptable for many users.

**5. Model-based properties.** Estimators that have been proposed have good properties under superpopulation models that generate them. This is important, particularly when the models are fit to explore relationships among the variables. Gelman rightly points out that an important issue is how to tell when we can have confidence in the regression coefficients from the model, particularly when many covariates are included.

**6. Robustness.** Holt and Smith (1979) emphasize robustness as one of the virtues of poststratification. Robustness is of course the big concern in any model-based weighting scheme, particularly when nonresponse or undercoverage occur since then one cannot check that the model holds for nonrespondents.

Some private survey organizations now take convenience samples and reweight the data in an attempt to generalize to the population (Schonlau, Fricker and Elliott, 2002). The accuracy of any population estimate from this method, such as the estimated percentage of people who think the government should provide health insurance for all children, then depends entirely on the model underlying the weighting scheme. If that model does not hold for individuals outside the sample, then the population estimates have unknown quality.

The tree of weights in Gelman's Figure 2 is a wonderful tool for studying the weights that result from various models. Figure 2 makes it clear that the big difference in the weight variability occurs in the example studied when education categories are added to the weighting model. Other trees could be drawn when the factors for weighting are considered in a different order, or when robust regression methods are used to estimate the parameters.

**7. Objectivity and perceived fairness of weighting procedure.** In most areas of statistics we strive for estimators with low MSE under reasonable models. In surveys, however, we want estimators that meet additional criteria. Because surveys are used for official statistics, those statistics should be acceptable by all participants in policy debates. Alexander (1994) emphasized that a survey statistician should be able to defend the choice of estimator to "politicians," defining "politician" as "anyone who is hoping to see your survey yield a particular result."

We should be able to defend the procedure used to develop weights to politicians and nonpoliticians alike. Any procedure used to construct weights will include subjective judgments—which variables to include in a model or how to construct and combine weighting adjustment cells. But persons who construct weights using models with the $y$ variable need to be especially careful that the models do not bias official statistics. This means careful attention to variable selection and influential observations, and exploration of a range of models.

Gelman's paper begins with the statement "Survey weighting is a mess." I do not think that survey weighting is a mess, but I do think that many people ask too much of the weights. For any weighting problem, one should begin by defining which of the possibly contradictory goals for the weights are desirable. Social scientists interested primarily in relationships among variables may value optimality under the model above all other features. I think that internal consistency of estimators and transparency of the weight construction method are essential for official statistics. Gelman's thought-provoking and informative paper made me think in a new way about weights, and I look forward to his future contributions to this discussion.

## ACKNOWLEDGMENTS

This research was supported in part by Grant SES-06-04373 from the National Science Foundation.